# PROBABILISTIC QUANTITATIVE PRECIPITATION FIELD FORECASTING USING A TWO-STAGE SPATIAL MODEL[1]

By Veronica J. Berrocal, Adrian E. Raftery
and Tilmann Gneiting

*Duke University, University of Washington and University of Washington*

Short-range forecasts of precipitation fields are needed in a wealth of agricultural, hydrological, ecological and other applications. Forecasts from numerical weather prediction models are often biased and do not provide uncertainty information. Here we present a postprocessing technique for such numerical forecasts that produces correlated probabilistic forecasts of precipitation accumulation at multiple sites simultaneously.

The statistical model is a spatial version of a two-stage model that represents the distribution of precipitation by a mixture of a point mass at zero and a Gamma density for the continuous distribution of precipitation accumulation. Spatial correlation is captured by assuming that two Gaussian processes drive precipitation occurrence and precipitation amount, respectively. The first process is latent and drives precipitation occurrence via a threshold. The second process explains the spatial correlation in precipitation accumulation. It is related to precipitation via a site-specific transformation function, so as to retain the marginal right-skewed distribution of precipitation while modeling spatial dependence. Both processes take into account the information contained in the numerical weather forecast and are modeled as stationary isotropic spatial processes with an exponential correlation function.

The two-stage spatial model was applied to 48-hour-ahead forecasts of daily precipitation accumulation over the Pacific Northwest

Received April 2008; revised August 2008.

[1]Supported in part by the DoD Multidisciplinary University Research Initiative (MURI) program administered by the Office of Naval Research under Grant N00014-01-10745, by the National Science Foundation under Awards ATM-0724721 and DMS-0706745, and by the Joint Ensemble Forecasting System (JEFS) under subcontract S06-47225 from the University Corporation for Atmospheric Research (UCAR).

in 2004. The predictive distributions from the two-stage spatial model were calibrated and sharp, and outperformed reference forecasts for spatially composite and areally averaged quantities.

**1. Introduction.** Due to its socioeconomic impact, precipitation is arguably the most important and most widely studied weather variable. Critical decisions in agriculture, hydrology, aviation, event planning and other areas depend on the presence or absence of precipitation, as well as precipitation accumulation. For such applications, reliable predictions of precipitation occurrence and precipitation amount are of great importance.

Operationally, short-range forecasts of precipitation are based on numerical weather prediction (NWP) models. However, despite an overall steady improvement in the quality of numerical weather predictions, forecasts of precipitation accumulation are still not as accurate and reliable as those of other meteorological variables [Applequist et al. (2002), Stensrud and Yussouf (2007)]. Furthermore, quantitative precipitation forecasts obtained from a single NWP model are deterministic, and thus do not convey any information about the uncertainty about the prediction, which is a shortcoming in weather-related decision-making [National Research Council (2006)]. One approach to incorporating uncertainty information into weather forecasting is via ensembles of numerical forecasts [Palmer (2002), Gneiting and Raftery (2005)]. While this is a major advance, the use of statistical postprocessing techniques for numerical forecasts remains essential.

Several methods have been developed to statistically postprocess numerical predictions of precipitation occurrence and produce probabilistic quantitative precipitation forecasts. They include linear regression [Glahn and Lowry (1972), Bermowitz (1975), Antolik (2000)], quantile regression [Bremnes (2004), Friederichs and Hense (2007)], logistic regression [Applequist et al. (2002), Hamill, Whitaker and Wei (2004)], neural networks [Koizumi (1999), Ramirez, de Campos Velho and Ferreira (2005)], binning techniques [Gahrs et al. (2003), Yussouf and Stensrud (2006)], hierarchical models based on climatic prior distributions [Krzysztofowicz and Maranzano (2006)], and two-stage models in which a Gamma density is employed to model precipitation accumulation [Wilks (1990), Hamill and Colucci (1998), Wilson, Burrows and Lanzinger (1999), Sloughter et al. (2007)].

All these methods treat forecast errors at different locations as spatially independent. This does not invalidate site-specific predictive distributions of precipitation. However, accounting for spatial correlation is critical for probabilistic forecasts of precipitation fields, or probabilistic forecasts of composite quantities, such as areally averaged precipitation accumulation, which are important in flood risk management and similar types of applications. Extended areas of high precipitation accumulation occur frequently in practice and incur much higher risk than would be expected under an assumption of spatial independence for the forecast errors.

In this paper we present a statistical method that postprocesses numerical forecasts of precipitation and yields calibrated probabilistic forecasts of daily precipitation accumulation at multiple sites simultaneously. Our approach builds on the two-stage model of Sloughter et al. (2007) and adds a spatial component to it, by using two spatial Gaussian processes driving, respectively, precipitation occurrence and precipitation accumulation. The first process is latent and results in a binary rain/no rain field; the second process drives precipitation amounts via an anamorphosis or transformation function [Chilès and Delfiner (1999), page 381]. The spatial dependence in the precipitation fields then derives from the spatial structure of the underlying Gaussian processes, which we model as stationary isotropic Gaussian processes equipped with exponential correlation structures.

At any individual site our model coincides with that of Sloughter et al. (2007), except that the latter has been developed for an ensemble of numerical forecasts, while our model uses a single numerical forecast only. Thus, at any individual site the predictive distribution of precipitation is a mixture of a point mass at zero and a Gamma distribution, with parameters that depend on the numerical forecast.

The paper is organized as follows. In Section 2 we give details of our statistical model, and we describe the numerical forecasts and precipitation data used in this study. In Section 3 we present results for probabilistic 48-hour-ahead forecasts of daily precipitation accumulation over the Pacific Northwest in the 2004 calendar year. We compare our method to competing prediction techniques, including an ensemble of NWP forecasts [University of Washington mesoscale ensemble; Grimit and Mass (2002), Eckel and Mass (2005)], the Bayesian model averaging technique of Sloughter et al. (2007) and versions of the power truncated normal model of Bardossy and Plate (1992). In Section 4 we review other statistical postprocessing approaches, and we discuss the limitations and some possible extensions of our method.

## 2. Data and methods.

2.1. *Numerical forecasts and precipitation data.*    To illustrate our method, we use observations and numerical predictions of daily (24-hour) precipitation accumulation during 2003 and 2004. The observations come from meteorological stations located in the Pacific Northwest, in a region centered on the states of Oregon and Washington, and are reported in whole multiples of one hundredth of an inch. Precipitation accumulations less than 0.01 inch were recorded as zeros.

The forecasts were provided by the Department of Atmospheric Sciences at the University of Washington. They are based on the MM5 [fifth-generation

Pennsylvania State University—National Center for Atmospheric Research Mesoscale Model; Grell, Dudhia and Stauffer (1995)] mesoscale numerical weather prediction (NWP) model, run with initial and boundary conditions provided by the United Kingdom Meteorological Office (UKMO). The NWP forecast was generated on a 12 km grid, at a prediction horizon of 48 hours, and bilinearly interpolated to observation sites. In total, our database consists of 109,996 observation/forecast pairs distributed over 560 days in the 2003 and 2004 calendar years. Note that the NWP forecast is one of the eight members of the University of Washington NWP ensemble [Eckel and Mass (2005)]. Our database contains the other ensemble members as well, but the UKMO member is considered the best.

Figure 1 shows forecasts and observations of daily precipitation accumulation valid for January 5, 2004. The gridded NWP forecast in panel (a) corresponds to the areally averaged precipitation accumulation over the 12 km grid cells. Panel (b) shows the NWP forecast at observation sites, obtained from the gridded forecast via bilinear interpolation. Panel (c) displays the observed precipitation accumulation. It is clear that the NWP model over-predicted precipitation accumulation. This wet bias was fairly typical. Over the 2003 and 2004 calendar years, the NWP model predicted precipitation accumulations larger than observed about 85% of the time, with a mean error of 4.45 hundredths of an inch. About 61% of the NWP forecasts indicated nonzero precipitation accumulations, while only 34% of the observations were nonzero. The other ensemble members showed similar wet biases.

Our goal in this paper is to develop a statistical method that corrects for the systematic bias present in the NWP forecast, yields calibrated predictive distributions for precipitation accumulation, and accounts for spatial correlation in the precipitation field.

2.2. *Spatial statistical model.* Several statistical models for precipitation occurrence and precipitation accumulation have been proposed in the literature. Stidd (1973), Bell (1987), Bardossy and Plate (1992), Hutchinson (1995), and Sansò and Guenni (1999, 2000, 2004) adapted a Tobit model [Tobin (1958), Chib (1992)] to precipitation accumulation, working with a latent Gaussian process that relates to precipitation via a power transformation and a truncation. The resulting power truncated normal (PTN) model offers a unified approach to precipitation modeling that allows both for a point mass at zero and a right-skewed distribution for precipitation accumulations greater than zero. However, it may not be flexible enough for our purposes, as we will see below.

Another approach to precipitation modeling uses two-stage models, which consider precipitation occurrence first, and then model nonzero precipitation accumulation conditional on its occurrence. Common choices for the distribution of nonzero precipitation accumulation include exponential densities [Todorovic and Woolhiser (1975)], mixtures of exponentials [Woolhiser and Pegram (1979), Foufoula-Georgiou and Lettenmaier (1987)] and Gamma densities [Stern and Coe (1984), Wilks (1989), Hamill and Colucci (1998), Wilson, Burrows and Lanzinger (1999), Sloughter et al. (2007)].

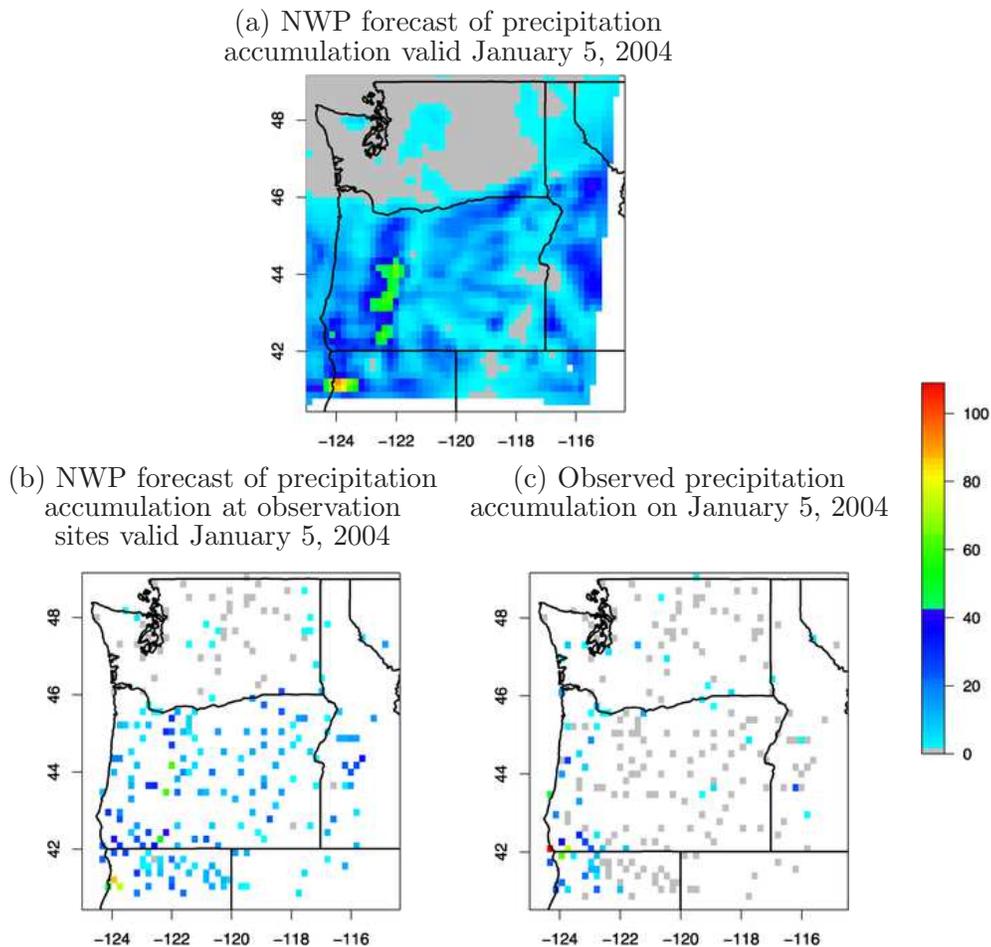

FIG. 1.   *NWP forecast and observations of daily precipitation accumulation valid for January 5, 2004, in hundredths of an inch, at a prediction horizon of 48 hours. The color grey is used to indicate no precipitation. (a) NWP forecast on a 12 km grid covering the Pacific Northwest. (b) NWP forecast interpolated to observation sites. (c) Observed precipitation accumulation.*

The spatial statistical model underlying our method is an extension of the two-stage model with a Gamma density for nonzero precipitation accumulation. From now on time is fixed, and so it is not explicitly included in the notation. Following Sloughter et al. (2007), we use the cube root of precipitation as the starting point of our model. Therefore, we denote by $Y(s)$ the cube root of the observed daily precipitation accumulation at the location $s$. We assume that there exists a latent Gaussian process $W(s)$ that drives precipitation occurrence. If $W(s)$ is less than or equal to zero, then there is no precipitation at the site; otherwise there is precipitation at $s$, that is,

$$Y(s) = 0 \qquad \text{if } W(s) \leq 0 \quad \text{and} \quad Y(s) > 0 \qquad \text{if } W(s) > 0.$$

We model the latent Gaussian process $W(s)$ as

$$W(s) = \mu(s) + \varepsilon(s), \tag{1}$$

where $\mu(s)$ is a spatial trend function that depends on the NWP forecast, and $\varepsilon(s)$ is a mean zero Gaussian spatial process. We follow Sloughter et al. (2007) in modeling the spatial trend as

$$\mu(s) = \gamma_0 + \gamma_1 \tilde{Y}(s) + \gamma_2 \mathbb{I}(s), \tag{2}$$

where $\tilde{Y}(s)$ is the cube root of the NWP forecast for the precipitation accumulation at $s$, and $\mathbb{I}(s)$ is an indicator variable equal to 1 if $\tilde{Y}(s) = 0$ and equal to 0 otherwise. At any individual site, this is simply a probit model for precipitation occurrence. The spatial Gaussian process $\varepsilon(s)$ has stationary and isotropic covariance function

$$\text{Cov}(\varepsilon(s), \varepsilon(s')) = \exp\left(-\frac{\|s - s'\|}{\rho}\right), \tag{3}$$

where $\|s - s'\|$ is the Euclidean distance between sites $s$ and $s'$. The parameter $\rho > 0$ is the range and specifies the rate at which the exponential correlation decays.

The second part of our model specifies the distribution of the cube root of precipitation accumulation given that there is precipitation, that is, conditionally on $Y(s)$ being greater than zero. At the marginal level, we model this conditional distribution by a Gamma distribution with site-specific parameters $\alpha_s$ and $\beta_s$, that is,

$$Y(s) \mid Y(s) > 0 \sim G_s = \text{Gamma}(\alpha_s, \beta_s). \tag{4}$$

Following Sloughter et al. (2007), we assume that the mean $\alpha_s \beta_s$ and the variance $\alpha_s \beta_s^2$ of the Gamma distribution in (4) depend on the NWP forecast. Specifically, we suppose that

$$\alpha_s \beta_s = \eta_0 + \eta_1 \tilde{Y}(s) + \eta_2 \mathbb{I}(s) \tag{5}$$

and

$$\alpha_s \beta_s^2 = \nu_0 + \nu_1 \tilde{Y}(s)^3, \tag{6}$$

where the parameters $\nu_0$ and $\nu_1$ are constrained to be nonnegative.

The model specification in (4), (5) and (6) refers to individual sites. However, our goal is to model precipitation at several sites simultaneously, so as to account for spatial dependence. Given the right-skewed distribution of precipitation accumulations, it is not possible to model the precipitation field directly using a spatial Gaussian process, so we consider a transformation approach. Let $G_s$ denote the Gamma distribution function in (4), and let $\Phi$ denote the standard normal distribution function. We assume that there exists a standardized Gaussian spatial process $Z(s)$ with covariance function

$$(7) \qquad \mathrm{Cov}(Z(s), Z(s')) = \exp\left(-\frac{\|s - s'\|}{r}\right),$$

such that, at each point $s$ at which $Y(s)$ is strictly positive,

$$(8) \qquad Y(s) = \Psi_s(Z(s)) = G_s^{-1} \circ \Phi(Z(s)),$$

where $\Psi_s = G_s^{-1} \circ \Phi$ is a spatially varying anamorphosis or transformation function [Chilès and Delfiner (1999), page 381]. The anamorphosis has the advantage of retaining the appropriate conditional distribution (4), while allowing us to model the spatial structure conveniently, using the Gaussian spatial process $Z(s)$. Note that (8) can be expressed as

$$(9) \qquad Z(s) = \Psi_s^{-1}(Y(s)) = \Phi^{-1} \circ G_s(Y(s)),$$

conditionally on $Y(s)$ being greater than zero. We refer to Barancourt, Creutin and Rivoirard (1992) and De Oliveira (2004) for additional discussion of this general type of random field model.

2.3. *Model fitting.* For forecasts on any given day, we estimate the parameters of the statistical model in Section 2.2 using observations and forecasts from a sliding training period made up of the most recent $M$ days for which they are available. We assume that the statistical relationships between the forecast and the observations are static during the training period, with any seasonal evolution captured by the rolling estimation window. Details on the choice of the length $M$ of the sliding training period will be given in Section 2.4.

In describing how we fit the model, we first explain how we go about estimating the parameters for precipitation occurrence, and then we present the procedure for precipitation accumulation.

In the model for precipitation occurrence, we estimate the trend parameters $\gamma_0, \gamma_1$ and $\gamma_2$ in (2) by a probit regression. The covariance parameter $\rho$ in (3) is estimated using the stochastic EM algorithm of Celeux and Diebolt (1985). The implementation requires simulation from a multivariate truncated normal distribution, for which we adopt the approach of Rodriguez-Yam, Davis and Scharf (2004).

We now turn to the model for precipitation accumulation. The anamorphosis function $\Psi_s$ that relates the precipitation field $Y(s)$ to the underlying Gaussian process $Z(s)$ is site-specific, because the Gamma distribution function $G_s$ in (4) varies spatially. To estimate the Gamma mean parameters $\eta_0, \eta_1$ and $\eta_2$ in (5), we fit a linear regression of the cube root of the nonzero observed precipitation accumulation on the cube root of the NWP forecast and the indicator of this forecast being equal to zero. The Gamma variance parameters $\nu_0$ and $\nu_1$ in (6) are estimated by numerically maximizing the marginal likelihood under the assumption of spatial and temporal independence of the forecast errors. To estimate the range parameter $r$ of the spatial Gaussian process $Z(s)$ in (7), we fix the other parameters at their estimates and maximize the marginal likelihood under the assumption of temporal independence. Calculating the Jacobian for the transformation (8), the likelihood for any given day in the training period is seen to be proportional to

$$(10) \qquad f_{Z(s_1),\ldots,Z(s_k)}(z_1,\ldots,z_k) \times \prod_{j=1}^{k} g_{s_j}(y_j)e^{z_j^2/2},$$

where $s_j$ is a site with observed precipitation accumulation greater than zero, $y_j > 0$ is the cube root of the precipitation amount, and $z_j = \Phi^{-1} \circ G_{s_j}(y_j)$, for $j = 1, \ldots, k$, with $k$ the number of sites with strictly positive observed precipitation accumulation on this day. The density $g_{s_j}$ is that of the Gamma distribution $G_{s_j}$, and $f$ is a zero mean Gaussian density that depends on the range parameter $r$ via (7). The full marginal likelihood is proportional to the product of (10) over the days in the training period and is optimized numerically.

2.4. *Choice of training period.* In principle, the longer the training period, the more data, and the more data, the better. On the other hand, a shorter training period allows changes in atmospheric regimes and the NWP model to be taken into account more promptly. To make an informed decision about the length of the training period, we consider the predictive performance of the two-stage spatial model at individual sites as a function of the length $M$ in days. To assess the quality of the predictive distributions for daily precipitation accumulation, we use the continuous ranked probability score [Matheson and Winkler (1976), Gneiting and Raftery (2007)], which is a strictly proper scoring rule for the evaluation of probabilistic forecasts of a univariate quantity. It is negatively oriented, that is, the lower the better, and is defined as

$$(11) \qquad \mathrm{crps}(F,x) = \int_{-\infty}^{\infty} (F(\xi) - \mathbb{I}\{x \le \xi\})^2 \, d\xi,$$

where $F$ is the predictive cumulative distribution function, $x$ is the realizing observation, and $\mathbb{I}$ is an indicator function. Gneiting and Raftery (2007) showed that (11) can be expressed equivalently as

$$(12) \qquad \text{crps}(F, x) = \text{E}_F |X - x| - \tfrac{1}{2} \text{E}_F |X - X'|,$$

where $X$ and $X'$ are independent random variables with common distribution $F$. In particular, if $F = F_{\text{ens}}$ is a forecast ensemble of size $m$ with members $x_1, \ldots, x_m$, then

$$(13) \qquad \text{crps}(F_{\text{ens}}, x) = \frac{1}{m} \sum_{i=1}^{m} |x_i - x| - \frac{1}{2m^2} \sum_{i=1}^{m} \sum_{j=1}^{m} |x_i - x_j|.$$

It is now immediate that the continuous ranked probability score is reported in the same unit as the forecast variable, and that it generalizes the absolute error, to which it reduces if $F$ is a point forecast.

Figure 2 shows the mean continuous ranked probability score as a function of the length $M$ of the rolling training period, where $M = 10, 15, 20, \ldots, 60$. The score is computed for predictive distributions of the original, nontransformed precipitation accumulation, so it has the unit of hundredths of an inch. It is temporally and spatially averaged over all predictive distributions at individual sites for the period March 9, 2003—March 8, 2004, at a prediction horizon of 48 hours, using the method described in the next section. The score improves (decreases) as the length $M$ of the rolling training period increases to 30 days, and thereafter does not change much. We therefore used a 30-day training period. A training period of length 30 days was also used by Sloughter et al. (2007), who applied a Bayesian model averaging (BMA) technique to this dataset. It is very possible that different choices would be best for other forecast lead times and other geographic regions.

2.5. *Generating forecasts.* Once the statistical model has been fitted, probabilistic forecasts of precipitation fields can be generated easily, by sampling from the underlying Gaussian processes $W(s)$ and $Z(s)$. We first simulate from the Gaussian process $W(s)$ that drives precipitation occurrence; then we generate realizations of the spatial Gaussian process $Z(s)$ at the sites $s$ where $W(s)$ is strictly positive. If $W(s) \leq 0$, then $Y(s) = 0$. If $W(s) > 0$, the realizations of $Z(s)$ are transformed into the cube root precipitation accumulation $Y(s)$ and the original precipitation accumulation $Y_0(s) = Y(s)^3$ using the site specific anamorphosis function (9).

We use this method to generate samples of any desired size from the joint predictive distribution of precipitation occurrence and precipitation accumulation on spatial grids. The simulation-based approach is a natural choice, because the model grid contains thousands of cells and it is not feasible to work with the resulting, very high-dimensional predictive distributions in closed form. The approach is illustrated in Figure 3, which shows two members of a statistical ensemble of precipitation field forecasts over the Pacific Northwest obtained with the two-stage spatial method. The forecasts are made at a 48 hour prediction horizon and valid January 5, 2004. The corresponding NWP forecast and the observed precipitation pattern are shown in panels (a) and (c) of Figure 1, respectively. The two-stage spatial post-processing method corrects for the wet bias present in the NWP model and provides a predictive distribution in the form of a statistical ensemble of precipitation fields, of any desired size.

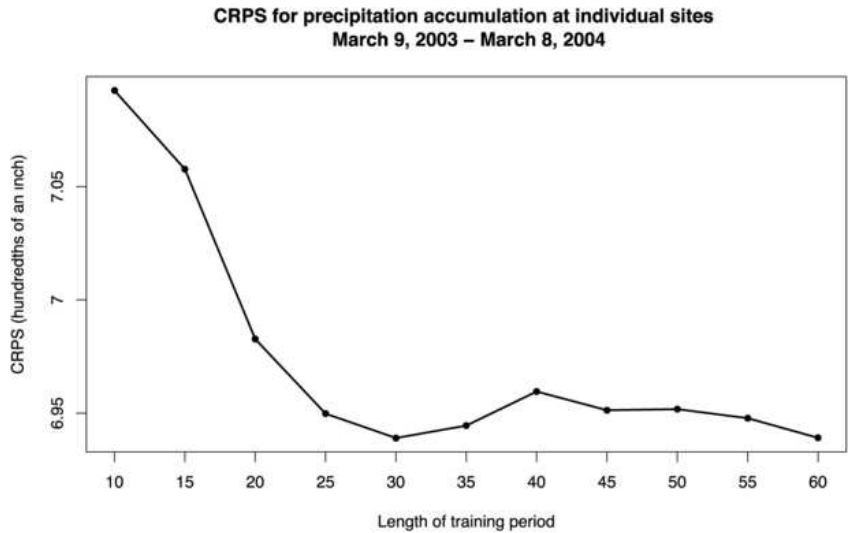

Fig. 2. *Mean continuous ranked probability score (CRPS) for probabilistic forecasts of daily precipitation accumulation at individual sites, for the period March 9, 2003–March 8, 2004, as a function of the length M of the sliding training period, in hundredths of an inch. The method used is the two-stage spatial technique.*

The spatial grid is of size approximately 10,000, so even simulation from the required multivariate normal distributions is not a straightforward task. For doing this, we use the circulant embedding technique [Wood and Chan (1994), Dietrich and Newsam (1997), Gneiting et al. (2006)] as implemented in the R package RandomFields [Schlather (2001)]. This is a very fast technique that can readily be used in real time.

For verification purposes, we need statistical forecast ensembles at observation sites, as opposed to the gridded forecasts in Figure 3. This can be done analogously, using NWP forecasts interpolated to observation sites as described in Section 2.1. However, the task is much easier computationally, since on average there were only 197 observation sites for precipitation accumulation in the Pacific Northwest on any given day.

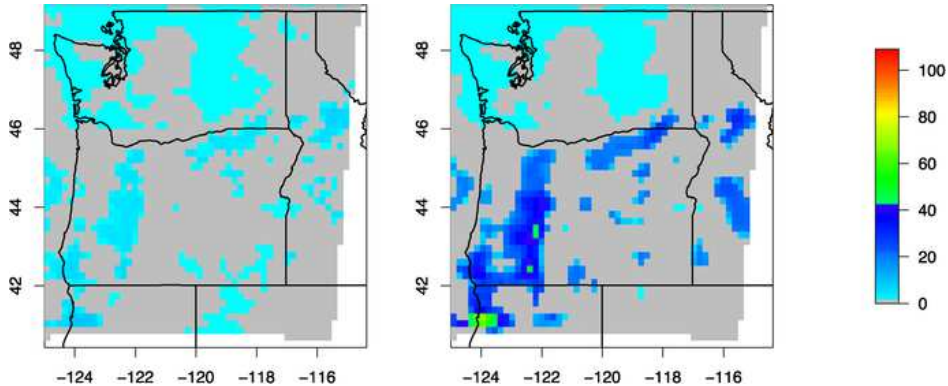

Fig. 3.  *Two members of a statistical forecast ensemble for daily precipitation accumulation over the Pacific Northwest valid for January 5, 2004 at a prediction horizon of 48 hours, using the two-stage spatial technique. Precipitation accumulation is indicated in hundredths of an inch, with the color grey representing no precipitation.*

## 3. Results.

3.1. *Reference forecasts.*   We now evaluate the out-of-sample predictive performance of our probabilistic forecasting method, to which we refer as the "two-stage spatial" method.

We used the two-stage spatial method to obtain forecasts of daily precipitation accumulation in the 2004 calendar year at observation sites over the Pacific Northwest and compared them to reference forecasts, as described below. All forecasts use a 48 hour prediction horizon and a sliding training period consisting of forecasts and observations for the most recent 30 days available, if applicable. Specifically, we consider the following types of forecasts:

(a)  As basic reference standard, we use "empirical climatology," that is, a static, temporally invariant predictive distribution that equals the empirical distribution of observed precipitation accumulation in the prior calendar year, 2003. Details are given below.

Table 1
*An overview of the forecast techniques used in the case study. See text for details*

| Forecasting technique | Gives predictive distribution | Uses NWP model | Uses NWP ensemble | Uses statistical postprocessing | Spatial modeling |
|---|---|---|---|---|---|
| Empirical climatology | yes | no | no | no | no |
| NWP | no | yes | no | no | no |
| NWP ensemble | yes | yes | yes | no | no |
| BMA | yes | yes | yes | yes | no |
| PTN | yes | yes | no | yes | yes |
| Two-stage spatial | yes | yes | no | yes | yes |

(b) The numerical forecast described in Section 2.1, namely, the UKMO member of the University of Washington NWP ensemble [Eckel and Mass (2005)], which we refer to as the "NWP" forecast.

(c) The full University of Washington NWP ensemble, which is a collection of eight numerical forecasts, each based on the MM5 NWP model, with initial and boundary conditions provided by eight distinct meteorological centers. We refer to this as the "NWP ensemble" forecast.

(d) The Bayesian model averaging (BMA) postprocessing technique of Sloughter et al. (2007) applied to the NWP ensemble in (c). The BMA predictive distribution is a mixture distribution, where each component is associated with an ensemble member and is based on a two-stage model that uses a Gamma density for precipitation accumulations greater than zero. The method ignores dependence of forecast errors between sites. We call this the "BMA" forecast.

(e) A postprocessing technique based on the NWP forecast in (b) and the power truncated normal (PTN) model of Bardossy and Plate (1992), in which a power transformed and truncated, stationary and isotropic spatial Gaussian process with mean structure similar to (5) and exponential correlation drives both precipitation occurrence and precipitation accumulation. The transformation power used here is $\gamma = 2$. We call this the "power truncated normal" or "PTN" method. See Berrocal (2007) for details.

(f) The PTN method in (e) with transformation power $\gamma = 2.33$, a value that is obtained by maximizing the marginal likelihood for this parameter.

(g) The "two-stage spatial" method described in this paper, which is a postprocessing technique based on the NWP forecast in (b).

Table 1 summarizes properties and characteristics of the various forecasting methods, which are listed roughly in order of increased complexity of the spatial modeling. The NWP forecast is deterministic; all the other methods are probabilistic, in that they provide predictive distributions. Among the probabilistic techniques, empirical climatology does not use any information from NWP models, as opposed to the others. The BMA method is a statistically postprocessed version of the NWP ensemble, but does not involve any spatial modeling. The PTN and two-stage spatial methods are built around a single NWP forecast, rather than an ensemble. They use statistical postprocessing to correct for biases and to generate predictive distributions, and employ spatial processes to account for spatial correlation in forecast errors.

In the remainder of this section we assess the predictive performance of these methods both marginally and jointly. For the marginal assessment, we evaluate forecasts of daily precipitation accumulation at individual sites. For the joint evaluation, we consider predictions of areally averaged precipitation accumulation, and forecasts of precipitation accumulation at several sites simultaneously. In our assessment, we are guided by the principle of maximizing the sharpness of the predictive distributions subject to calibration [Gneiting, Balabdaoui and Raftery (2007)]. In other words, we aim at predictive distributions that are as concentrated as possible, while being statistically consistent with the observations. To provide summary measures of predictive performance that address calibration and sharpness simultaneously, we use strictly proper scoring rules, such as the continuous ranked probability score, the Brier score and the energy score [Gneiting and Raftery (2007)].

3.2. *Verification results for precipitation accumulation at individual sites.* We now present verification results for probabilistic forecasts of daily precipitation accumulation at individual sites in the Pacific Northwest in the 2004 calendar year. Numerical forecasts and observations were available for a total of 249 days in 2004. All results and scores are spatially and temporally aggregated, comprising a total of 66,663 individual forecast cases at a prediction horizon of 48 hours. Our basic reference standard is empirical climatology, here taken to be the static, spatially and temporally invariant predictive distribution that equals the empirical distribution of observed precipitation accumulation, when aggregated over the 2003 calendar year and the Pacific Northwest.

Table 2 shows summary measures of predictive performance, including the mean absolute error (MAE) and mean continuous ranked probability score (CRPS) for precipitation accumulation, and the mean Brier score (BS) for precipitation occurrence. The absolute error is a performance measure for a deterministic forecast, here taken to be the median of the predictive distribution. The continuous ranked probability score (11) is a proper scoring rule for a probabilistic forecast of a scalar quantity; for a deterministic forecast, it reduces to the absolute error. The Brier score or quadratic score [Brier (1950)] for a probability forecast of a binary event is defined as

$$\mathrm{bs}(f, o) = (f - o)^2,$$

where $f$ is the forecast probability for the event and $o$ equals 1 if the event occurs and 0 otherwise. As the representation (12) shows, the continuous ranked probability score for a predictive distribution equals the integral over the Brier score for the induced probability forecasts at all real-valued thresholds $\xi$. The entry in the table refers to precipitation occurrence, that is, the threshold $\xi = 0$.

The table shows that the statistically postprocessed forecasts (BMA, PTN and two-stage spatial method) outperformed the others. The BMA forecast had slightly lower scores than the two-stage spatial and PTN methods; this is not surprising, given that it is based on the full NWP ensemble rather than a single member only. The superiority of the two-stage spatial method

|                         | MAE  | CRPS | BS    |
|-------------------------|------|------|-------|
| Empirical climatology   | 7.71 | 7.19 | 0.222 |
| NWP                     | 9.55 | 9.55 | 0.325 |
| NWP ensemble            | 8.46 | 6.76 | 0.271 |
| BMA                     | 6.68 | 5.02 | 0.141 |
| PTN ($\gamma = 2$)      | 7.17 | 5.63 | 0.164 |
| PTN ($\gamma = 2.33$)   | 6.99 | 5.53 | 0.148 |
| Two-stage spatial       | 6.73 | 5.12 | 0.148 |

over the PTN technique may stem from a lack of flexibility of the latter, as it depends on a power transform and attempts to accommodate precipitation occurrence and precipitation accumulation using a single latent spatial process.

To assess the calibration of the predictive distributions, we use verification rank histograms [Anderson (1996), Talagrand, Vautard and Strauss (1997), Hamill and Colucci (1997), Hamill (2001)] and probability integral transform (PIT) histograms [Diebold, Gunther and Tay (1998), Gneiting, Balabdaoui and Raftery (2007)]. Verification rank histograms are used for ensemble forecasts when the number of members $m$ is small. For each forecast case, the rank of the verifying observation is tallied within the combined set of $m + 1$ values given by the ensemble members and the observation. If the ensemble members and the observation are exchangeable, the verification rank follows a discrete uniform distribution over the set $\{1, 2, \ldots, m+1\}$. Thus, under the assumption of exchangeability and over a large number of forecast cases, the verification rank histogram is expected to be statistically uniform. Similarly, the PIT histogram displays the PIT value, that is, the value that the predictive cumulative distribution function attains at the observation. If the observation is a random draw from the forecast distribution, the PIT value is uniformly distributed, and over a large number of forecast events, we expect the PIT histogram to be uniform. Deviations from uniformity can be interpreted diagnostically in terms of dispersion errors and biases [Diebold, Gunther and Tay (1998), Hamill (2001), Gneiting, Balabdaoui and Raftery (2007)].

Predictive distributions for quantitative precipitation have point masses at zero, so to retain uniformity under the null assumption, we need to randomize. We first consider forecasting methods that produce a NWP ensemble. In situations in which the observation and one or more ensemble mem-bers equal zero, we draw a verification rank from the set $\{1, \ldots, m_0 + 1\}$, where $m_0$ is the number of ensemble members equal to zero.

In the case of the PIT histogram, in instances in which the observation equals zero, a PIT value is obtained by drawing a random number from a uniform distribution between 0 and the predicted probability of precipitation. With these modifications, verification rank and PIT values remain uniformly distributed under the corresponding null assumptions.

Figure 4 shows verification rank histograms and PIT histograms for the various types of probabilistic forecasts. The NWP ensemble consists of eight members, that is, the verification ranks range from 1 to 9. The ensemble is underdispersed and has a wet bias, so the observations tend to overpopulate the lowest rank, which is seen in the rank histogram. The other techniques are considerably better calibrated, with the two-stage spatial method showing the most uniform PIT histogram. The histograms for the PTN technique indicate that observations of precipitation accumulation have heavier tails than can be modeled by a power transformed normal distribution.

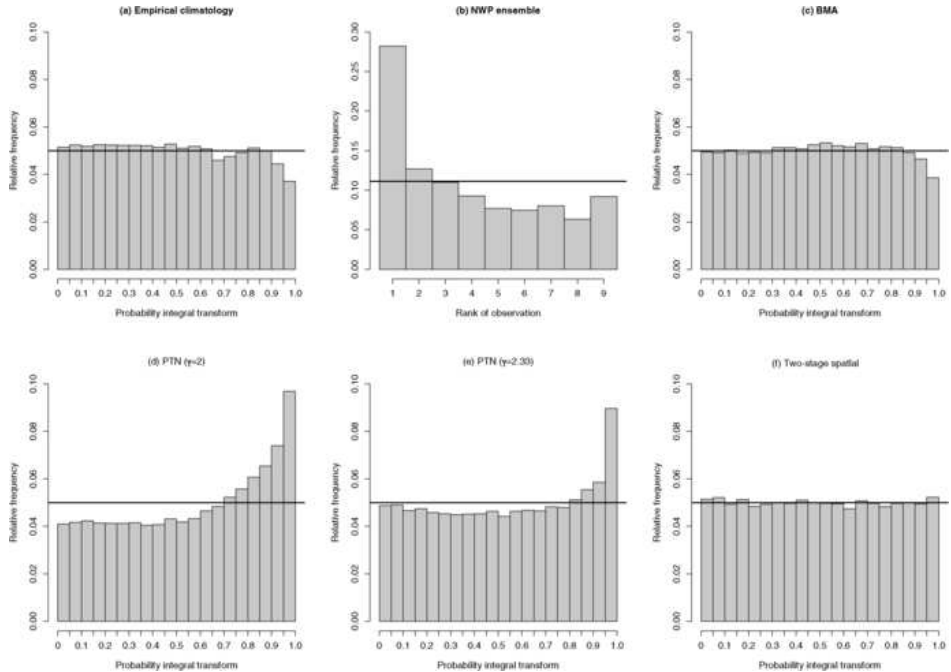

FIG. 4. *Verification rank and probability integral transform (PIT) histograms for probabilistic forecasts of daily precipitation accumulation at individual sites, temporally and spatially aggregated over the 2004 calendar year and the Pacific Northwest.* (a) *Empirical climatology.* (b) *NWP ensemble.* (c) *BMA method.* (d) *PTN method with $\gamma = 2$.* (e) *PTN method with $\gamma = 2.33$.* (f) *Two-stage spatial technique.*

**Reliability diagram for probability of precipitation**

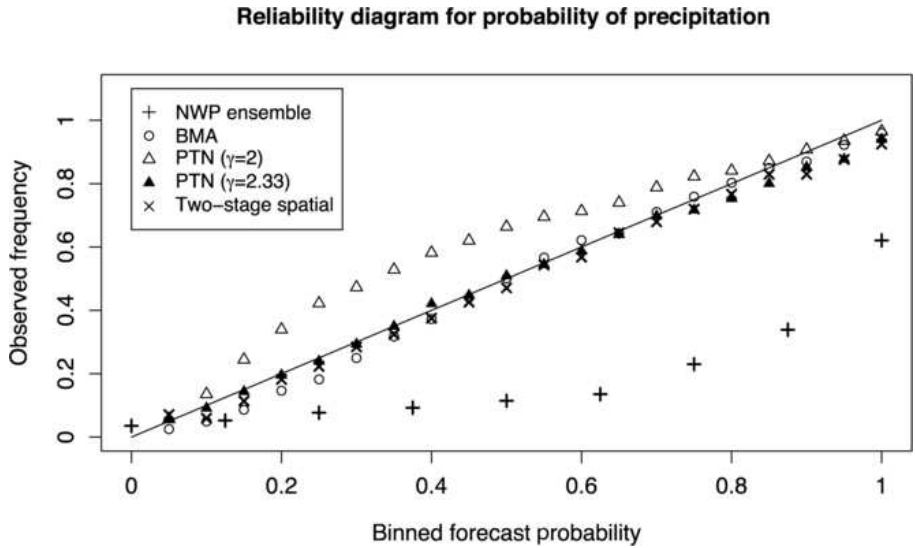

Fig. 5. *Reliability diagram for probability forecasts of precipitation occurrence at individual sites, for the various types of forecasts, temporally and spatially aggregated over the 2004 calendar year and the Pacific Northwest.*

We complete this section by assessing the reliability of the induced probability forecasts for the occurrence of precipitation. The reliability diagram in Figure 5 shows the empirically observed frequency of precipitation occurrence as a function of the binned forecast probability. For a calibrated forecast, we expect the graph to be close to the diagonal. Due to its wet bias, the NWP ensemble tends to overpredict precipitation occurrence, which results in a reliability curve below the diagonal. The BMA method, the PTN technique with $\gamma = 2.33$ and the two-stage spatial method were reliable. Overall, the BMA and two-stage spatial methods performed best.

3.3. *Verification results for areally averaged precipitation accumulation.* When predicting spatially composite quantities, it can be critically important that spatial correlation be taken into account. One such quantity, which is important in hydrological and agricultural applications, is total or average precipitation over an area, such as a river catchment. Probabilistic forecasts of the average precipitation accumulation over a region $\mathcal{A}$ with area $|\mathcal{A}|$ can be derived easily using the two-stage spatial method. Let $Y_0(\mathcal{A})$ denote the average precipitation accumulation over $\mathcal{A}$, and write $Y_0(s) = Y(s)^3$ for the original, nontransformed precipitation accumulation at the site $s \in \mathcal{A}$, expressed in terms of the cube root accumulation $Y(s)$. Then

$$Y_0(\mathcal{A}) = \frac{1}{|\mathcal{A}|} \int_{\mathcal{A}} Y_0(s)\, ds,$$

which can be approximated by the composite quantity

$$(14) \qquad \bar{Y}_0 = \frac{1}{J}\sum_{j=1}^{J} Y_0(s_j) = \frac{1}{J}\sum_{j=1}^{J} Y(s_j)^3,$$

where $s_1, \ldots, s_J$ are sites located within $\mathcal{A}$. The two-stage spatial method allows us to sample from the predictive distribution of $\bar{Y}_0$ as follows:

(i) Generate a realization of the latent Gaussian process $W(s)$ at the sites $s_1, \ldots, s_J$ using (1), (2) and (3).

(ii) Generate a realization of the spatial Gaussian process $Z(s)$ at the sites $s_j$ at which $W(s_j) > 0$ using (7).

(iii) If $W(s_j) \leq 0$, let $Y(s_j) = 0$. If $W(s_j) > 0$, find $Y(s_j)$ using (9) and the site specific Gamma parameters in (5) and (6).

(iv) Find a realization of the composite quantity $\bar{Y}_0$ using (14).

We applied this method to generate probabilistic forecasts of areally averaged daily precipitation accumulation over the Upper Columbia River basin in 2004 using the two-stage spatial method, and compared to reference techniques. The Columbia River basin is a 259,000-square-mile basin that spans seven states (Oregon, Washington, Idaho, Montana, Nevada, Wyoming and Utah) and one Canadian province (British Columbia). It is the most hydroelectrically developed river system in the world, with more than 400 dams and a generating capacity of 21 million kilowatts.

Here, we consider only the upper part of the Columbia River basin that lies within the state of Washington. Fifteen of the 441 meteorological stations in our data base are located in this area. On 212 days in 2004, two or more of these stations reported daily precipitation accumulation, so we consider the composite quantity (14), where $J$ may vary from day to day. The minimum, median and maximum of $J$ among the 212 forecast cases were 2, 10 and 14, respectively. To obtain a predictive distribution for the composite quantity $\bar{Y}_0$ with the two-stage spatial method, we repeated steps (i) through (iv) to obtain a sample of size 10,000. For verification purposes, this can be handled as a continuous predictive distribution, and we do so in the following. The reference forecasts are treated analogously.

Table 3 shows summary measures of predictive performance. The PTN and two-stage spatial methods, which invoke statistical postprocessing and model spatial structure, outperformed the other techniques. The two-stage spatial method performed best, showing both the lowest MAE and the lowest CRPS.

Figure 6 shows verification rank and PIT histograms for the probabilistic forecasts. The rank histogram for the NWP ensemble is U-shaped and left-skewed, as a result of its underdispersion and wet bias. The PIT histogram for the BMA technique is also U-shaped; its underdispersion stems from the fact that it does not take account of spatial dependence. A similar pattern is seen for empirical climatology, hinting at interannual variability that cannot be captured by a one-year record. The PIT histograms for the PTN method point at the aforementioned tail issues. The empirical distribution of areally averaged precipitation accumulation has a heavier tail than the PTN method allows, so PIT values close to 1 appear too often. The PIT histogram for the two-stage spatial method is the most uniform.

TABLE 3
*Mean absolute error (MAE) and mean continuous ranked probability score (CRPS) for forecasts of areally averaged daily precipitation accumulation over the Upper Columbia River basin in 2004*

|  | **MAE** | **CRPS** |
| --- | --- | --- |
| Empirical climatology | 5.78 | 4.72 |
| NWP | 7.76 | 7.76 |
| NWP ensemble | 7.99 | 6.20 |
| BMA | 5.31 | 4.01 |
| PTN ($\gamma = 2$) | 5.04 | 3.74 |
| PTN ($\gamma = 2.33$) | 5.05 | 3.77 |
| Two-stage spatial | 4.90 | 3.63 |

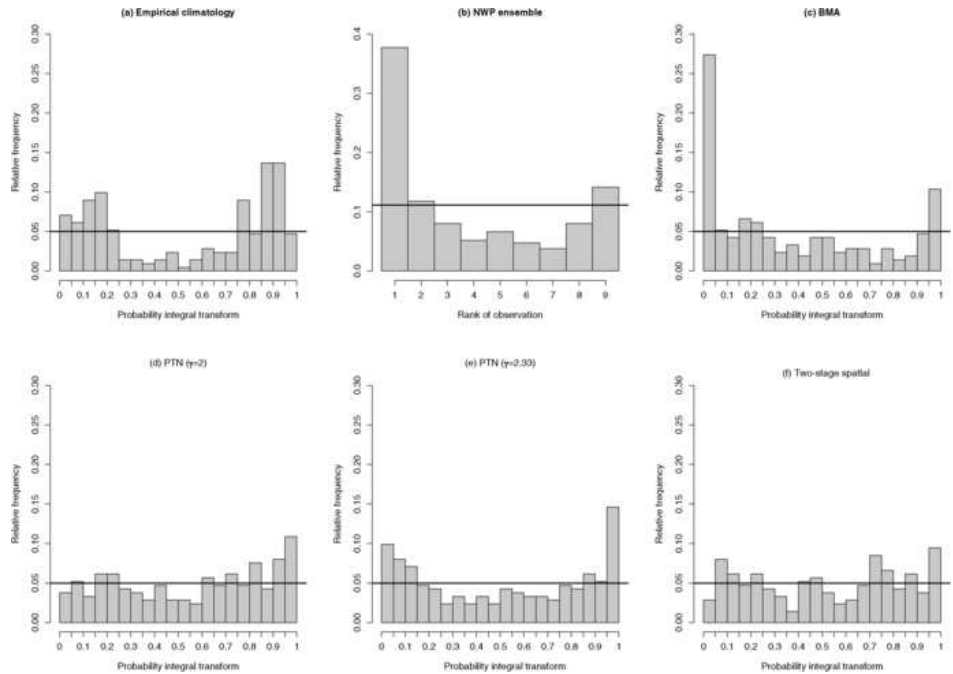

FIG. 6. *Verification rank and probability integral transform (PIT) histograms for probabilistic forecasts of areally averaged daily precipitation accumulation over the Upper Columbia River basin in 2004.* (a) *Empirical climatology.* (b) *NWP ensemble.* (c) *BMA technique.* (d) *PTN method with $\gamma = 2$.* (e) *PTN method with $\gamma = 2.33$.* (f) *Two-stage spatial technique.*

3.4. *Spatial verification.* To assess further whether the forecasting methods capture spatial correlation, we consider multivariate probabilistic forecasts of daily precipitation accumulation at several sites simultaneously. In the experiment reported here, we selected the four stations in the Upper Columbia River basin that had the most observations in 2004, namely, Brown Mountain Orchard, Gold Mountain, Nespelem and Teepee Seed Orchard, which have a median inter-station distance of 43 miles. Observations of daily precipitation accumulation at these four stations simultaneously were available on 141 days in the 2004 calendar year.

For these 141 days, we generated four-dimensional probabilistic forecasts of precipitation accumulation at these sites, using the same techniques, 48 hour prediction horizon and 30 day sliding training period as before. In the case of empirical climatology, we used the four-dimensional empirical distribution of observed precipitation accumulation at the four sites in 2003. For the other methods, we generated statistical ensembles from the joint predictive distribution of precipitation accumulation. For the BMA method, this four-dimensional distribution has independent components; for the PTN and two-stage spatial methods, the components are correlated.

Given that the predictive distributions are for a four-dimensional, vector-valued quantity, we need to adapt our verification methods [Gneiting et al. (2008)]. For a combined assessment of sharpness and calibration, we use the energy score. Specifically, if $F$ is the predictive distribution for a vector-valued quantity and $\mathbf{x}$ materializes, the energy score is defined as

$$\text{es}(F, \mathbf{x}) = \text{E}_F \|\mathbf{X} - \mathbf{x}\| - \tfrac{1}{2} \text{E}_F \|\mathbf{X} - \mathbf{X}'\|, \tag{15}$$

where $\| \cdot \|$ denotes the Euclidean norm and $\mathbf{X}$ and $\mathbf{X}'$ are independent random vectors with common distribution $F$. Note that (15) is a proper scoring rule that is a direct multivariate generalization of the continuous ranked probability score in the kernel representation (12). In particular, if $F = F_{\text{ens}}$ is an ensemble forecast with vector-valued members $\mathbf{x}_1, \ldots, \mathbf{x}_m$, then

$$\text{es}(F_{\text{ens}}, \mathbf{x}) = \frac{1}{m} \sum_{i=1}^{m} \|\mathbf{x}_i - \mathbf{x}\| - \frac{1}{2m^2} \sum_{i=1}^{m} \sum_{j=1}^{m} \|\mathbf{x}_i - \mathbf{x}_j\|,$$

which is a multivariate generalization of (13). Like the continuous ranked probability score, the energy score is negatively oriented.

*Mean energy score for ensemble forecasts of daily precipitation accumulation at four sites in the Upper Columbia River basin simultaneously, in 2004*

|  | Energy score |
|---|---|
| Empirical climatology | 12.96 |
| NWP | 20.72 |
| NWP ensemble | 15.08 |
| BMA | 10.49 |
| PTN ($\gamma = 2$) | 10.57 |
| PTN ($\gamma = 2.33$) | 10.90 |
| Two-stage spatial | 10.45 |

To assess calibration for ensemble forecasts of multivariate weather quantities, we use the minimum spanning tree (MST) rank histogram [Smith and Hansen (2004), Wilks (2004)]. If the ensemble has $m$ members, the MST rank is found by tallying the length of the MST that connects the $m$ ensemble members within the combined set of the $m + 1$ lengths of the ensemble-only MST and the $m$ MSTs obtained by substituting the observation for each of the ensemble members. If the ensemble members and the observation are exchangeable, these lengths are also exchangeable. Therefore, for a calibrated forecast technique and over a large number of forecast events, we expect the MST rank histogram to be statistically uniform. For an underdispersed ensemble, the lowest ranks are overpopulated.

Verification results for the four-dimensional probabilistic forecasts of precipitation accumulation at Brown Mountain Orchard, Gold Mountain, Nespelem and Teepee Seed Orchard are shown in Table 4 and Figure 7. The two-stage spatial method has the lowest energy score, with the PTN techniques and, perhaps surprisingly, the BMA method being close competitors. The MST rank histogram for the NWP ensemble is based on the $m = 8$ members of the University of Washington ensemble and attests to its underdispersion, which is typical for unprocessed NWP ensembles. The MST rank histograms for the other methods are computed from statistical ensembles with $m = 19$ members. They are nearly uniform for the BMA, PTN and two-stage spatial techniques.

**4. Discussion.** We have presented a statistical method for obtaining probabilistic forecasts of precipitation fields from a numerical forecast. The method builds on the two-stage model of Sloughter et al. (2007) developed for precipitation forecasts at individual sites, and extends it by accounting for spatial correlation. At any individual site, the distribution of precipitation is modeled by a mixture of a point mass at zero and a Gamma distribution for precipitation accumulations greater than zero. The spatial dependence between precipitation at different sites is captured by introducing two spatial Gaussian processes, that drive, respectively, precipitation occurrence and precipitation accumulation. The latter process is linked to precipitation via a site specific transformation function. This allows us to retain the marginal Gamma distribution while conveniently modeling the spatial correlation using techniques for Gaussian random fields. The method entails an implicit downscaling, in which NWP forecasts on a 12 km grid scale are statistically corrected to apply to observation sites.

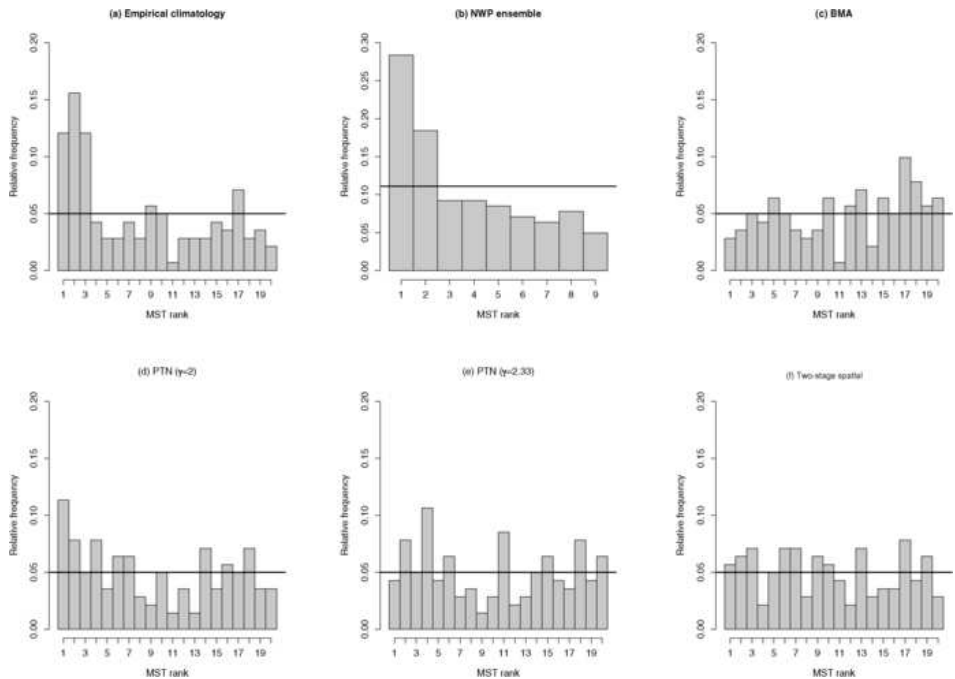

Fig. 7. *Minimum spanning tree (MST) rank histograms for ensemble forecasts of daily precipitation accumulation at four sites in the Upper Columbia River basin simultaneously, in 2004.* (a) *Empirical climatology.* (b) *NWP ensemble.* (c) *BMA technique.* (d) *PTN method with $\gamma = 2$.* (e) *PTN method with $\gamma = 2.33$.* (f) *Two-stage spatial technique.*

In a case study on probabilistic forecasts of daily precipitation accumulation over the Pacific Northwest in 2004, the two-stage spatial model captured the spatial dependence in precipitation fields. It resulted in predictive distributions which generally were calibrated and outperformed reference forecasts. The increased flexibility of the two-stage spatial model over the BMA method stems from the fact that it accounts for spatial correlation, while the BMA method does not. The power truncated normal (PTN) technique also accounts for spatial dependence; however, it is less flexible than the two-stage spatial method, since it uses a power transformation and relies on a single Gaussian process to accommodate both precipitation occurrence and precipitation accumulation.

Typically, statistical postprocessing methods for precipitation accumulation operate site by site [Applequist et al. (2002)]. However, a number of methods to generate correlated probabilistic forecasts of precipitation accumulation at several sites simultaneously have been proposed. Possibly the most prevalent approach is the aforementioned PTN technique, which has been adapted by Bardossy and Plate (1992) and Sansò and Guenni (2004) to honor information from NWP models. The method of Seo et al. (2000) is a downscaling technique that generates ensembles of precipitation fields at a finer spatial resolution than the original model grid. Kim and Mallick (2004) explored the use of skew-Gaussian random fields in precipitation forecasting. Herr and Krzysztofowicz (2005) proposed a bivariate statistical model for precipitation at two locations that uses a two-stage approach with a meta-Gaussian distribution that represents nonzero precipitation accumulation. Unlike ours, the method is restricted to two sites and does not exploit the information in NWP models.

There are various ways in which the two-stage spatial method could be expanded. The spatial processes that account for the spatial correlation in precipitation occurrence and precipitation accumulation are modeled as stationary and isotropic Gaussian processes with an exponential correlation function. More general covariance structures such as the Matérn covariance function [Stein (1999), Guttorp and Gneiting (2006)] could be employed. It would also be possible to adopt the fully Bayesian approach described by De Oliveira, Kedem and Short (1997). However, this would be much more computationally intense and might be impractical in real time, where fast implementation is vital.

Finally, the two-stage spatial method is built around a single member of the University of Washington NWP ensemble [Eckel and Mass (2005)]. It seems feasible, though technically difficult, to account for the flow-dependent uncertainty information contained in the NWP ensemble by combining our method with the full Bayesian model averaging (BMA) framework of Sloughter et al. (2007). This would be similar to the way in which Berrocal, Raftery and Gneiting (2007) combined the geostatistical model of Gel, Raftery and Gneiting (2004) and the BMA technique of Raftery et al. (2005) to provide probabilistic forecasts of temperature fields, but would be considerably more complex due to the non-Gaussian character of precipitation fields. With the continued development of NWP ensemble systems, the combined method remains a challenge for future work; at present, its marginal benefits are likely to be incremental.

**Acknowledgments.** We are grateful to Jeff Baars, Chris Fraley, Clifford F. Mass and J. McLean Sloughter for discussions and providing code and data.

V. J. BERROCAL
DEPARTMENT OF STATISTICAL SCIENCE
DUKE UNIVERSITY
DURHAM, NORTH CAROLINA 27705
USA
E-MAIL: vjb2@stat.duke.edu

A. E. RAFTERY
T. GNEITING
DEPARTMENT OF STATISTICS
UNIVERSITY OF WASHINGTON
SEATTLE, WASHINGTON 98195-4322
USA
E-MAIL: raftery@u.washington.edu
        tilmann@stat.washington.edu